\documentclass{PoS}

\title{Non-planar structure of analytical QCD predictions for 
the Gottfried sum rule}

\ShortTitle{Non-planar  QCD predictions and the  Gottfried sum rule}

\author{\speaker{A.L.Kataev}\thanks{Supported in part by RFBR Grants 
N 05-01-00992, N 06-02-16659}\\
Institute for Nuclear Research of the Russian  Academy of Sciences
,117312  Moscow, Russia
               \\
        E-mail: \email{kataev@ms2.inr.ac.ru}}
\abstract{
 It is stressed that  within large $N_c$-expansion  
{\bf analytical calculations} of the 
$(\alpha_s/\pi)^2$ QCD  contributions to the valence part of the 
Gottfried sum rule  for $F_2$ structure function 
of  charged leptons-nucleon deep-inelastic scattering   reveals the existence  
of the $O(1/N_c^2)$ {\bf non-planar} corrections   only   
and the    disappearance of the {\bf planar} $O(N_c^0)$ perturbative terms.
The relation between Gottfried and Adler 
sum rule for neutino-nucleon DIS  is established and the proposal 
that the differebce between  corresponding QCD
corrections to higher non-singlet moments in charged-lepton and neutrino 
DIS are {\bf suppressed by $1/N_c^2$} is made. 
The possible consequence of the cancellation of {\bf perturbative 
planar} graphs in the  
considered perturbative series, namely the 
existence of  light-quark flavour asymmetry 
$\overline{u}(x)<\overline{d}(x)$, is mentioned.
The effect of the similar origin, i.e. the  relation 
of  {\bf  ``light-by-light-type''-type  structure}  
in  DIS characteristics at the $\alpha_s^3$-level to the  
generation  of   light quark-antiquark asymmetry,  
is also  commented.}

\FullConference{XI International Workshop on Advanced Computing 
and Analysis Techniques in Physics Research\\
		 April 23 2007\\
		 Amsterdam, the Netherlands}

\begin{document}

\section{Introduction}
Modern calculation physics has several important 
features, which may  push ahead this part of science. 
To our point of view among them are: 
\begin{enumerate}  
\item the requirement of {\bf reliability} 
of  the complicated numerical or analytical 
calculations, related to  the{\bf  elimination of 
bugs} in the computer codes and final results.  
It   seems to be obvious to realize this requirement,  but sometimes 
not simple  (see e.g. \cite{Hatton});
\item the requirement  {\bf to avoid} shortcomings of definite 
numerical calculations, which may manifest themselves in the  
{\bf non-proper rounding off errors} in  the  
numerical results (see e.g.\cite{Hatton}, \cite{Morhac}); 
\item the  interest   to {\bf  phenomenological applications} 
of  various perturbative results and the development 
of different  resummations approaches   
of  perturbative series (see e.g. \cite{Beneke:1998ui},
\cite{Shirkov:2006gv});
\item the appearance of strong intention  to go beyond straightforward  
theoretical calculations  and the desire to    find 
new mathematical  \cite{Broadhurst:1998rz}  and  physical   
\cite{Broadhurst:2004jx}
properties, 
related to {\bf the differences  in the   topological 
 structure} 
of various  multiloop Feynman diagram, which define 
{\bf special  
 analytical structures} of perturbative series under investigation. 
\end{enumerate}
I will try to demonstrate that continuing at present  
QCD calculations 
of characteristics of deep-inelastic scattering (DIS) satisfy 
all  requirements 
 mentioned above. In particular, I will concentrate myself 
on discussions of possible 
relations of the concrete  analytical structures in  the concrete  
 results of 
perturbative QCD calculations to the existence in nature of  
definitely  
speaking non-perturbative effects, namely  
light quarks  
flavour asymmetry 
$\overline{u}(x)<\overline{d}(x)$ (see Ref.\cite{Broadhurst:2004jx})
Related phenomenological manifestations  of this effects will be also 
considered. Presented in Ref.\cite{Broadhurst:2004jx}  conjecture  about 
existence of  differences between   
QCD corrections to the non-singlet (NS) moments of charged leptons - nucleon
and neutrino-nucleon DIS, which manifest themselves in the 
large $N_c$-limit, proposed in   Ref.\cite{'tHooft:1973jz}, 
will be described.The perturbative indications to the generation of     
$s(x)-\overline{s}(x)$ strange quark  asymmetry (see Ref.\cite{Catani:2004nc} 
will be mentioned as well. 

\section{Definitions of some  perturbative series  in DIS  } 

Let me follow normalization conditions of the related 
perturbative series,  
used  in the work of Ref. \cite{Kataev:2001kk}  devoted to  
the detailed fits of the experimental 
data for the NS $xF_3$ SF of   $\nu N$ DIS, provided by CCFR collaboration
\cite{Seligman:1997mc}.
Consider  DGLAP equation in the NS  case
\begin{equation}
\label{DGLAP}
Q^2\frac{d}{dQ^2}F^{NS}(x,Q^2)=\frac{1}{2}\int_x^{1}\frac{dz}{z}
\bigg[P_{NS}(z,a_s)+\beta(a_s)\frac{\partial{\rm ln}C_{NS}(z,a_s)}
{\partial a_s}\bigg]F^{NS}\bigg(\frac{x}{z},Q^2\bigg)
\end{equation}
where $C_{NS}(z,a_s)$ and $P_{NS}(z,a_s)$ are the NS combinations of 
the   coefficient functions and splitting functions, 
$F^{NS}$- NS combinations of   $F_2^{l,\nu}$ structure functions (SFs) 
of charged leptons ( $l$) or $\nu$ - nucleon DIS.
The 
QCD $\beta$-function is defined as 
\begin{equation}
\mu\frac{\partial a_s}{\partial\mu}=\beta(a_s)=-2\sum_{n\geq 0}
\beta_n a_s^{n+2}~~.
\end{equation}
where  
$a_s=\alpha_s/(4\pi)$ and the first two scheme-independent coefficients 
are normalized as 
\begin{eqnarray}
\label{beta0}
\beta_0&=&\bigg(\frac{11}{3}C_A-\frac{2}{3}N_F\bigg) \\ 
\label{beta1}
\beta_1&=&\bigg(\frac{34}{3}C_A^2-2C_FN_F-\frac{10}{3}C_AN_F\bigg)
\end{eqnarray}
The general expressions for the Casimir 
operators of the $SU(N_c)$ group of colour are   
$C_A=N_c$,  $C_{F}=(N_c^2-1)/(2N_c)$, while $N_F$ is the number of 
quarks flavours.
The  series for the coefficient  functions and  
splitting  functions of  Eq.(\ref{DGLAP}) read
\begin{eqnarray}
\label{CF}
C_{NS}(z,a_s)&=&\sum_{n\geq 0} C_n^{NS}(z)a_s^n \\
\label{NS}
P_{NS}(z,a_s)&=&\sum_{n\geq 0 } P_{NS}^{(n)}(z)a_s^{n+1}
\end{eqnarray}
The corresponding non-expanded in $a_s$ NS  splitting functions 
we will be interested in 
have the form, given 
in the  work of Ref.\cite{Moch:2004pa}  
\begin{equation}
{P}_{NS}^{\pm}(x)={P}_{qq}^v(x)\pm {P}_{q\overline{q}}^v(x)
\end{equation} 
\begin{eqnarray}
{P}_{q_{i}q_{k}}& =& {P}_{\overline{q}_{i} \overline{q}_{k}}
= \delta_{ik}{P}_{qq}^{v}+P_{qq}^{s} \\
{P}_{q_{i}\overline{q}_{k}}& =& {P}_{\overline{q}_{i} q_{k}}
= \delta_{ik}{P}_{q\overline{q}}^{v}+P_{q\overline{q}}^{s}~~.
\end{eqnarray}
where the next-to-next-to-leading order (NNLO)  $a_s^3$ corrections 
to the NS combinations of  valence 
( ${P}_{qq}^v$,~ ${P}_{q\overline{q}}^{v}$)   
and sea ($P_{qq}^{s}(x)$,~$P_{q\overline{q}}^{s})$ contributions  
were derived  analytically in  \cite{Moch:2004pa}.

These results  are related to the coefficients of the 
anomalous dimensions of NS moments 
 through the following Mellin 
transformation   
\begin{equation}
\label{ad}
\gamma_{NS}^{N}(a_s)=
\sum_{n\geq 0} \gamma_{n}^{N} a_s^{n+1}  \\
= -2\int_0^1 dx\bigg[ P_{NS}(x,a_s)\bigg]{x}^{N-1} 
= -2 \int_0^1 dx \bigg[\sum_{n\geq 0}P_{NS}^{(n)}(x)\bigg]a_s^{n+1}~{x}^{N-1}
\end{equation}
In the case of N=2,4,6,8  
Mellin NS moments of  the charged lepton-nucleon DIS SF $F_2$
they were previously 
calculated analytically at the NNLO  
in Ref. \cite{Larin:1993vu},  N=10  
NNLO  results were obtained in  Ref.\cite{Larin:1996wd},  while for 
N=12,14 they are known from the calculations of Ref.\cite{Retey:2000nq}.
In the case of 
 N=16 NS moment of $F_2$ SF  the 
direct calculation of the 
$O(a_s^3)$ approximation for  $\gamma_{NS}^{N}(a_s)$, performed  in Ref. 
\cite{Blumlein:2004xt}, helped to check  
the three-loop results for  $P_{NS}(x,a_s)$ \cite{Moch:2004pa} by 
verifying coincidence of anomalous dimension of  
N=16   moment, calculated  in Ref.\cite{Blumlein:2004xt}, with the 
result obtained from the expression  of Ref.\cite{Moch:2004pa} 
using Mellin transformation   in  Eq.(\ref{ad}).

It should be stressed, that 
among  the first phenomenological applications 
of the results of Refs.\cite{Larin:1993vu},\cite{Larin:1996wd} were 
the NNLO DIS fits of Refs.\cite{Parente:1994bf},\cite{Kataev:1996vu},
\cite{Kataev:1997nc},\cite{Santiago:1999pr},\cite{Kataev:1999bp},  
\cite{Kulagin:2000yw}, where the works of Refs,\cite{Parente:1994bf},
\cite{Santiago:1999pr} were devoted to the analysis of $F_2$ SF data. 
The works of Refs.\cite{Kataev:1996vu},\cite{Kataev:1997nc}, 
\cite{Kataev:1999bp},\cite{Kulagin:2000yw} were aimed to    
the   ``approximate'' NNLO fits of 
CCFR $xF_3$ data for $\nu N$ DIS,  
which were essentially based 
on  the results  of rather complicated and, to my point of view, 
distinguished calculations of the NNLO corrections $C_2^{xF_3}(x)$  
to the coefficient 
function of the DGLAP equation for  $xF_3$ SF of $\nu N$ DIS, 
performed  by  van Neerven and Zijlstra \cite{Zijlstra:1992kj}.
The first confirmation of this result came from the coincidence 
of its first Mellin  moment with the NLO analytical approximation for the 
Gross-Llewellyn Smith sum rule, evaluated by Gorishny and Larin   
\cite{Gorishnii:1985xm}. More detailed cross-checks were done  in 
Ref.\cite{Moch:1999eb} using different methods.

The concrete theoretical arguments, that for moderate and high N 
the NNLO approximations of the   NS anomalous dimensions for $F_2$ SF
will be rather closed to the NNLO approximations of NS anomalous 
dimensions for $xF_3$ SF were used  in Ref. \cite{Kataev:1996vu}.
Theoretical studies of these arguments, which 
in part are based  on the proposal to use extrapolation procedure 
from even  to 
odd  values of N  \cite{Parente:1994bf}, are on the agenda 
(careful checks  of the validity of the 
extrapolations  procedures  was done  in Ref.\cite{Kataev:2001kk} and 
in the resent work of Ref.\cite{Moch:2007gx}). 

The ``approximate'' feature of the  NNLO fits of $xF_3$ data 
 pushed a-head  analytical calculations of 
the NNLO approximations for  
NS anomalous dimensions and coefficient functions of $xF_3$ SF of  odd  
Mellin moments with N=1,3,5,7,9,11,13 \cite{Retey:2000nq}
(the   $N=1$  NNLO  results of Ref.\cite{Retey:2000nq}  
coincide with the NNLO calculations  
for the Gross-Llewellyn Smith sum rule \cite{Larin:1991tj}).

Let us now  return to the definition of the coefficient function 
in  Eq.(\ref{CF}) in the case of NS combinations of $F_2$ SF.  
The NNLO  term of order $a_s^2$ 
can be divided in two parts
\begin{equation}
C_2^{NS,F_2}(x)=C^{(2),(+)}(x)\pm C^{(2),(-)}(x)
\end{equation}
where the sign {\bf plus}  in between  two terms 
 corresponds to the NS contribution 
to the coefficient function of the  $F_2$ SF of charged lepton-nucleon DIS,
while  the sign {\bf minus}
appears, when the coefficient function of the NS contributions to   $F_2$ SF 
of  neutrino-nucleon  DIS  is considered.
The expressions for the terms $C^{(2),(+)}(x)$  and $C^{(2),(-)}(x)$,
were calculated  in Ref.\cite{van Neerven:1991nn} and confirmed in 
Ref.\cite{Moch:1999eb}. Thus, all discussions presented above, 
are demonstrating that the {\bf  first requirement},
mentioned  in the Introduction, i.e. the {\bf  reliability}
of the results obtained,   is fulfilled in the case of the 
complicated NNLO calculations of DIS characteristics.
The {\bf third} requirement, and in particular, its first 
part, namely the 
{\bf applications of new perturbative results} for   {\bf the phenomenological 
studies}, is also fulfilled in the case of DIS characteristics.

Indeed. 
the appearance of the calculations of Ref.\cite{Retey:2000nq}
gave the possibility to perform  more careful analysis 
of CCFR  $xF_3$ structure functions 
data, which came  from the studies of $\nu N$ DIS process 
at  Tevatron. These phenomenological  fits  were 
made in  Refs.\cite{Santiago:2001mh},
\cite{Kataev:2001kk},\cite{Maxwell:2002mt},\cite{Khorramian:2006wg},
\cite{Brooks:2006wh}) using different 
theoretical approaches.  Note, however, 
that after these works the number of questions 
still remain for possible 
future studies.
For example,  
in spite of the fact, that the NNLO theoretical inputs of various  
fitting codes 
of analyzing DIS
experimental data ( i.e. the ones, 
based on application of the    DGLAP equation and on the reconstruction of 
structure functions from their moments using   
Bernstein and Jacobi polynomials)    
are coordinated, the outputs sometimes differ drastically.

Among the concrete   questions   is   the existence  of 
``small'' NNLO value  $\alpha_s(M_Z)=0,1142\pm 0.0008$
obtained by Bernstein polynomial analysis of the CCFR $xF_3$ measurements 
\cite{Khorramian:2006wg}, which is agreement  with   
``small''
NNLO result $\alpha_s(M_Z)=0.1141\pm 0,0014$ of the   combined 
NNLO  fits of various DIS data \cite{Alekhin:2005gq} (other ``small''
 NNLO values were recently  discussed in  
Ref.\cite{Blumlein:2007dk}), but is not consistent  
with  the ``world average'' result 
$\alpha_s(M_z)=0.1189\pm 0.0010$ \cite{Bethke:2006ac}. 
It is also worth to mention, that, 
that the ``small'' result of Ref.\cite{Khorramian:2006wg} 
does  not also agree with other NNLO applications of the Bernstein 
polynomial technique  for studies of the same data CCFR $xF_3$ data, which 
give the following results  
$\alpha_s(M_Z)=0.1166\pm 0.0013$ \cite{Santiago:2001mh},
$\alpha_s(M_Z)=0.1196^{+0.0027}_{-0.0031}$
\cite{Maxwell:2002mt} and  $\alpha_s(M_Z)=0.1189 \pm 0.0019$ 
\cite{Brooks:2006wh}.
It will be of interest to trace  the origin of these   disagreements
of applications of the same data and same 
theoretical method in the different works on the subject.
Moreover, ``small'' NNLO   $\alpha_s(M_Z)$ values,  extracted from various 
characteristics of DIS,
do not agree 
with the the recent NNLO result $\alpha_s(M_Z)=0.119 \pm 0.002~(exp)
\pm 0.003~(theory)$, obtained within the   updated set of NNLO parton 
distributions of UK group \cite{Martin:2007bv}. It should be stressed, 
that this value  is in   perfect agreement with the one, obtained  from the  
NNLO fits of $xF_3$ CCFR data using Jacobi polynomial technique,  
namely $\alpha_s(M_Z)=0.119\pm 0.005~(exp)
\pm~0.003 ~(theory)$ \cite{Kataev:2001kk}, where experimental error 
is dominated by systematic uncertainties of CCFR $\nu N$ scattering DIS 
data. We think that more careful coordination  of 
theoretical and experimental uncertainties of different 
 NNLO  determinations 
of $\alpha_s(M_Z)$ from the characteristics of DIS data will help to 
resolve existing disagreements. 
   
\section{Adler and Gottfried  sum rules: the definitions}  
  
Two other sections are  following  the original presentation 
at ACAT07 Workshop  
of the materials of the work,  made in collaboration 
with Broadhurst and Maxwell (see Ref.\cite{Broadhurst:2004jx} ).
Some additional considerations \cite{Kataev:2004wv},   
which are  supporting  the results of 
Ref.\cite{Broadhurst:2004jx} will be also commented. 

The  isospin Adler sum rule \cite{Adler:1965ty} is  
 expressed through the  structure function $F_2^{\nu N}$ of neutrino-nucleon 
 DIS as 
 \noindent
 \begin{equation}
 I_A=\int_0^1\frac{dx}{x}\bigg[F_2^{\nu p}(x,Q^2)-F_2^{\nu n}(x.Q^2)\bigg]
 =4I_3=2~~~.
 \label{eq:Adler}
 \end{equation}
In terms of parton distributions Eq. (\ref{eq:Adler}) takes the following form 
 \noindent
 \begin{equation}
 I_A=2\int_0^1dx[u_v(x)-d_v(x)]=2
 \label{eq:valence}
 \end{equation}
 where $u_v(x)=u(x)-\overline{u}(x)$ and $d_v(x)=d(x)-\overline{d}(x)$ are 
 the valence parton distributions. 
It was shown, that 
$I_A$ 
receives neither perturbative nor non-perturbative $(1/Q^2)$-corrections 
\cite{Dokshitzer:1995qm}. In view of this the Adler sum rule  is 
$Q^2$ independent and demonstrates the consequence of  
the property  of scaling \cite{Bjorken:1968dy}.
This property (or  so called ``automodelling'' behavior
of  structure functions) was rigorously proved  
 in the case of charged lepton-nucleon DIS    
 \cite{BVT} with application of general principles of local quantum field 
theory, described e.g. in the classical  text-book \cite{BSh}.
It should be stressed, that the validity of this sum rule is supported 
by the  existing neutrino--nucleon
DIS data, which show no significant $Q^2$ variation in the range
$2~{\rm GeV}^2\leq Q^2\leq 30~{\rm GeV}^2$ and give~\cite{Allasia:1985hw}
\begin{equation}
I_A^{\rm exp}=2.02\pm 0.40\;.
\end{equation}
Though the error-bars are quite large, the precision could
in principle may be improved by analyzing data of other  
$\nu N$ DIS experiments in more detail.
  
 However, it is known, that in QCD scaling is violated. 
The sources of its violation are related to the asymptotic freedom effects
and non-perturbative contributions. Both types of these effects manifest 
themselves (though is some 
puzzling way) in the analog 
of the Adler sum rule, namely in the Gottfried sum rule. 
It can be  defined as the first $N=1$ NS
Mellin moment of the difference of $F_2$ SFs of DIS of charged leptons on 
proton and neutron, namely   
\noindent
\begin{equation}
I_G^{v}=\int_0^1\frac{dx}{x}\bigg[F_2^{l p}(x,Q^2)-F_2^{l n}(x.Q^2)\bigg]
 =\frac{1}{3}\int_0^1 dx \bigg(u_v(x,Q^2)-d_v(x,Q^2)\bigg) =\frac{1}{3}
 \label{eq:Got}
 \end{equation}
 The definition of Eq.~(\ref{eq:Got}) is presented in the case of 
assumption  accepted  previously  that the sea quarks are  
 flavour-independent.
 It corresponds to the condition $\overline{u}(x,Q^2)=\overline{d}(x,Q^2)$, 
 accepted in the early works on the subjects. 
However, due to the appearance  of  experimental data 
for the muon--nucleon DIS, Drell-Yan process  and semi-inclusive DIS 
 we know at present that this condition is violated and 
  $\overline{u}(x,Q^2)<\overline{d}(x,Q^2)$ ( for reviews see, 
 e.g. \cite{Kumano:1997cy}-\cite{ALK}). Therefore, the definition of 
 the  Gottfried sum rule was  modified as:  
 \noindent
 \begin{equation}
 I_G=\int_0^1\frac{dx}{x}\bigg[F_2^{l p}(x,Q^2)-F_2^{l n}(x.Q^2)\bigg]
 = I_G^v+\frac{2}{3}
 \int_0^1 dx \bigg(\overline{u}(x,Q^2)-\overline{d}(x,Q^2)\bigg)~~,
 \label{eq:Got1}
 \end{equation}
 where the last term is related 
to the manifestation of isospin-breaking effects in the Dirac sea.
Without it  the valence part of the Gottfried 
sum rule of  Eq.~(\ref{eq:Got}) contradicts the existing result of 
the  
analysis  of muon--nucleon DIS data, performed  by the NMC collaboration,
namely \cite{Arneodo:1994sh}:
\begin{equation}
\label{NMC}
I_G(Q^2=4~\rm{GeV}^2)=0.235\pm 0.026~.
\end{equation}
Note, however, that  
the error  of this experimental result may be over two times  larger due to 
more careful treatment of the uncertainty from the small 
$x$ region \cite{Abbate:2005ct}. In spite of this, definite contradictions 
of theory with the quark-parton result still exist \cite{Arneodo:1994sh}
and it is worth to study the contributions of the perturbative and 
non-perturbative corrections to the valence part $I_G^v$.

\section{ Large $N_c$-expansion 
of the  Gottfried  sum rule and its relation to the   Adler sum rule.} 

Let us now discuss the results of calculations of perturbative 
QCD corrections to the valence part of the Gottfried sum rule,
following the  considerations of  Ref. \cite{Broadhurst:2004jx}
and using the ideology of large-$N_c$ limit, proposed in 
Ref.\cite{'tHooft:1973jz}. 
The solution of the renormalization group equation for the valence 
 contribution $I_G^v$ to the Gottfried sum rule has the following 
 form
 \noindent
 \begin{equation}
 I_G^v= A(\alpha_s)C^{(l)}(\alpha_s)
 \label{eq:RG}
 \end{equation}
 with the anomalous-dimension term
 \noindent 
\begin{eqnarray}
A(a_s)
&=&exp\bigg[-\int_\delta^{a_s(Q^2)}\frac{\gamma^{(l)~N=1}_{NS}(x)}{\beta(x)}dx
\bigg]= 1+\frac{1}{8}\frac{\gamma_1^{(N=1)}}{\beta_0}\bigg(\frac
 {\alpha_s}{\pi}\bigg)\\ \nonumber
&+&\frac{1}{64}\bigg(\frac{1}{2}\frac{(\gamma_1^{(N=1)})^2}
 {\beta_0}-\frac{\gamma_1^{(N=1)}\beta_1}{\beta_0^2}+\frac{\gamma_2^{(N=1)}}
 {\beta_0}\bigg)\bigg(\frac{\alpha_s}{\pi}\bigg)^2+O(\alpha_s^3)~~~.
 \end{eqnarray} 
The anomalous dimension function of $N=1$ moment  is defined as  
\begin{equation}
\label{AD}
\gamma^{(l)~N=1}_{NS}=\sum_{n\geq 0} \gamma_i^{(l)~N=1}a_s^{i+1} 
\end{equation}
The first 
coefficient $\gamma_0^{(l)~N=1}$  
is identically equal to 
{\bf zero} \cite{Gross:1973ju}. 
The leading correction 
to  Eq.(\ref{AD}) comes from the scheme-independent two-loop contribution 
\cite{Ross:1978xk} 
\begin{equation}
\gamma_1^{(l)~N=1}=-4(C_F^2-C_AC_F/2)[13+8\zeta(3)-12\zeta(2)]
\label{gamma1}
\end{equation}
 confirmed  
later on in Ref.\cite{Curci:1980uw},  
Notice the appearance of the distinctive {\bf non-planar colour factor} 
$(C_F^2-C_AC_F/2)=O(N_c^0)$, which exhibits $O(1/N_c^2)$ suppression at 
large-$N_c$, in comparison with 
the individual weights of planar two-loop diagrams, 
namely  $C_F^2$ and $C_FC_A$, that are  canceling    
in the expression for  $\gamma_1^{(l)~N=1}$. 
 
The three-loop coefficients of anomalous dimensions of N-th moments  
moments for the NS combinations of structure 
functions $F_2$ of   neutrino  - nucleon and charged lepton-nucleon   DIS  
are related to the Mellin moments from the 
splitting functions as  \cite{Moch:2004pa}
\begin{eqnarray}
\label{nu}
\gamma_{2}^{(\nu)~N}&=&2 \int_{0}^{1}{dx}\bigg[{P}_{NS}^{(2)-}(x)+{P}_{NS}^
{(2)+}(x)\bigg]
{x}^{N-1}  \\
\label{l}
\gamma_{2}^{(l)N}&=&2 \int_{0}^{1}{dx}\bigg[{P}_{NS}^{(2)-}(x)-{P}_{NS}^{(2)+}
(x)\bigg]
{x}^{N-1}~~.
\end{eqnarray}
Taking into account the nullification of anomalous dimension 
function for the Adler sum rule, one can derive 
the useful  identity 
\begin{equation}
\gamma_{2}^{(l)~N=1}=-2 \int_{0}^{1} dx {P}_{NS}^{(2)+}(x)~~.
\end{equation}
Integrating   the results  Ref.\cite{Moch:2004pa} and taking into 
account  mathematical analytical 
procedures of Ref.\cite{Borwein:1999js}, in Ref.\cite{Broadhurst:2004jx}  
the  analytical 
expression for $\gamma_{2}^{(l)~N=1}$ was obtained: 
\begin{eqnarray}
\label{gamma2}
\gamma_2^{(l)~N=1}&=&
(C_F^2-C_AC_F/2)\bigg\{
C_F\bigg[
290-248\zeta(2)
+656\zeta(3)
-1488\zeta(4)+832\zeta(5)
+192\zeta(2)\zeta(3)
\bigg]
\nonumber
\\&&{}
+C_A\bigg[
\frac{1081}{9}+\frac{980}{3}\zeta(2)-
\frac{12856}{9}\zeta(3)
+\frac{4232}{3}\zeta(4)
-448\zeta(5)
-192\zeta(2)\zeta(3)
\bigg]
\nonumber
\\&&{}
+N_F\bigg[
-\frac{304}{9}-\frac{176}{3}\zeta(2)+\frac{1792}{9}\zeta(3)
-\frac{272}{3}\zeta(4)\bigg]\bigg\} \\
\nonumber
&\approx&161.713785 - 2.429260\,N_F~~~~.
\end{eqnarray}

Notice the appearance in $\gamma_2^{(l)~N=1}$ of {\bf three non-planar 
factors},
namely $C_F^2(C_F-C_A/2)$ and  $C_FC_A(C_F-C_A/2)$ and $C_F(C_F-C_A/2)N_F$.
These results  are generalizing the observation 
of {\bf non-planarity} of  $\gamma_1^{(l)~N=1}$-term of anomalous dimension 
function to three-loops and 
may be considered as the first non-obvious argument in favor 
of the correctness   of definite results of Ref. \cite{Moch:2004pa}. 

Let us turn to 
the coefficient function of the valence contribution to the Gottfried 
sum rule $I_G^{v}$. It  has the following form 
\begin{equation}
{C}^{(l)}(a_{s})=
\frac{1}{3}\left[1+
\sum_{n=1}^{\infty}{C}_{n}^{(l)N=1}
{\left(\frac{\alpha_s}{\pi}\right)}^{n}\right]
\end{equation}
Like in the case of the Adler sum rule, the first 
coefficient ${C}_{1}^{(l)N=1}$ is  equal to  {\bf zero}
(see e.g. Ref.\cite{Curci:1980uw}). 
The the second coefficients to the N-th  Mellin moments 
of the NS combinations  of structure 
functions $F_2$  of   charged lepton-nucleon and neutrino-nucleon    DIS  
are related to the Mellin moments from Eq.(2.11) 
as 
\begin{eqnarray}
\label{GS}
{C}_{2}^{(l)N}& =&
3\int_{0}^{1}{dx}[{C}^{(2),(+)}(x)+{C}^{(2),(-)}(x)]{x}^{N-1}~~\\
{C}_{2}^{(\nu)N} &=&
\frac{1}{2}\int_{0}^{1}{dx}[{C}^{(2),(+)}(x)-{C}^{(2),(-)}(x)]{x}^{N-1}
\end{eqnarray} 
In the case of N=1 the explicit {\bf numerical  integration} 
of Eq.(\ref{GS}) gave the following result \cite{Kataev:2003en} 
\begin{equation}
C_2^{(l)~N=1}=3.695C_F^2-1.847C_AC_F
\label{num}
\end{equation}
where  ${C}^{(2),(\pm)}(x)$  in Eq.(\ref{GS}) were taken 
from Ref.\cite{van Neerven:1991nn}
with the  choice of $\overline{\rm{MS}}$
renormalization scale $\mu^2$=$Q^2$.
Note, that non fulfillment of the {\bf second requirement},  mentioned in 
the Introduction, namely the  {\bf non-proper rounding off errors 
of numerical integrations of Ref.\cite{Kataev:2003en}}
did not allow to     
 realize  that 
Eq. (\ref{num}) has the {\bf same non-planar} structure as 
the expression 
for $\gamma_1^{(l)~N=1}$. This fact 
was demonstrated in Ref. \cite{Broadhurst:2004jx}, where the following 
analytical result for $C_2^{(l)~ N=1}$ was obtained
\begin{equation}
C_2^{(l)N=1}=(C_F^2-C_AC_F/2)\bigg[-\frac{141}{32}+\frac{21}{4}\zeta(2)
-\frac{45}{4}\zeta(3)+12\zeta(4)\bigg]~~~~.
\label{anal}
\end{equation}
with the help of  the Mellin representation 
\begin{equation}
C_{2}^{(l)~N=1}= 2\times 3 \int_0^1dx [{C}^{(2),(-)}(x)] 
\end{equation}
which follows from Eqs.(\ref{nu}), (\ref{l}) and the requirement 
of absence of QCD corrections to the Adler sum rule.

Next, taking the large $N_c$ limit for Eq.(4.1), which presumes that   
the corrections to the coefficients of QCD  
$\beta$-function of Eq.(\ref{beta0}) and of Eq.(\ref{beta1}) 
take the following form 
\begin{eqnarray}
\label{betaN}
\beta_0&=& \frac{11}{3}N_c \bigg(1+O(\frac{N_F}{N_c})\bigg) \\ 
\label{beta1N}
\beta_1&=&\frac{34}{3}N_c^2\bigg(1+O(\frac{N_F}{N_c})\bigg)
\end{eqnarray}
and $\alpha_s/(\pi N_c) \sim 1/(N_c^2 ~ln(Q^2/\Lambda^2))$,  
we find that  Gottfried and Adler sum rules are rfelated as
\begin{equation}
\label{rel}
I_G^{v}=\frac{2}{3}I_A\bigg(1+O(\frac{1}{N_c^2})\bigg)~~~~~.
\label{relation}
\end{equation} 
This relation  is scheme-independent. Indeed, the transformation 
of the $\alpha_s$-corrections in  
Eq. (\ref{eq:RG}) to another 
$MS$-like scheme,   can be done with  the help of the shift 
\noindent
\begin{equation}
\frac{\alpha_s(Q^2)}{\pi}=\frac{\alpha^{'}_s(Q^2)}{\pi}+ \beta_0\Delta 
\bigg(\frac{\alpha_s^{'}(Q^2)}{\pi}\bigg)^2
\label{trans}
\end{equation}
 where $\Delta$ is the concrete $N_c$-independent  number, 
which is defined by the 
logarithm  from the ratio of regularization scales $\mu^2_{\overline{MS}}$
and $\mu^2_{MS-like}$. Thus, the general $MS$-like scheme 
expression for the coefficient  $C_2^{(l) N=1}$
takes the following form
\noindent
\begin{equation}
C_{2~~ MS-like}^{(l)N=1}= C_{2~~\overline{MS}}^{(l)N=1}+\gamma_1^{N=1}\Delta
\label{shift}
\end{equation}
where both  $C_{2~~\overline{MS}}^{(l)N=1}$ and $\gamma_1^{N=1}$ contain
the same {\bf non-planar} group weight $C_F(C_F-C_A/2)$ \cite{Kataev:2004wv}. 
\newpage
Keeping in mind the intriguing relation between the  
valence part of the  Gottfried sum rule and the  Adler sum rule, one may 
make a prediction, that  the difference between 
NS moments of $F_2$ in charged-lepton and neutrino scattering will 
continue to satisfy the property of {\bf non-planar} suppression    
\cite{Broadhurst:2004jx}. For example the ratio
\begin{equation}
\label{ratio}
R_2^N\equiv{C_2^{(l)N}-6C_2^{(\nu)N}\over C_2^{(l)N}+6C_2^{(\nu)N}}
={\int_0^1{\rm d}x\,C^{(2),(-)}(x)x^{N-1}\over
\int_0^1{\rm d}x\,C^{(2),(+)}(x)x^{N-1}}
\end{equation}
takes the value $R_2^{N=1}=1$ at $N=1$, but  
for $N=2$, one cane  obtained from Ref.\cite{Moch:1999eb}
\begin{equation}
\label{ratio2}
R_2^{N=2}=-\frac{0.50593104}{5.4183241N_c^2 -4N_cN_F - 8.4480127}
\end{equation}
which is negative and small in magnitude at large $N_c$, and is continuing to 
decrease rapidly with increasing of N (see discussions in 
Ref.\cite{Broadhurst:2004jx}). So, we expect that the ratio
\begin{equation}
\frac{6C_n^{(\nu)N}}{C_n^{(l)N}}=1+O(\frac{1}{N_c^2})
\end{equation}
is valid not only for 2 loops, 
but in higher orders of perturbation theory as well
\cite{Broadhurst:2004jx}.

It should be stressed, that the results and   conjectures of   
Ref.\cite{Broadhurst:2004jx}, which were  discussed in this talk,   
already received several confirmations.
Indeed, in Ref.\cite{Kotikov:2005gr} the analytical expressions of 
Eq.(\ref{gamma2}) and Eq.(\ref{anal}) were  reproduced with the help of 
calculations, which are based on different theoretical technique.
One more argument in favor of  
careful  study of  
the topological structure of analytical perturbative series 
for characteristics of DIS  \cite{Broadhurst:2004jx} came 
from the recent analytical calculations of the NS moments 
of $\nu N$ DIS both at the $a_s^2$ and $a_s^3$  level \cite{Rogal},
and in the confirmation of the special role of 
{\bf non-planar-type} colour structures in higher corrections of the 
quantities considered. Thus, two works, presented at
this workshop, satisfy the {\bf fourth} requirement 
of importance of analytical 
calculations  mentioned,  in the Introduction, namely the importance 
of careful study of {\bf topological structute} of perturbative QCD series. 

\section{Large $N_c$-limit: where are the effects of $O(N_C^0)$ 
corrections to the Gottfried sum?}

First of all it it worth to stress that {\bf non-planar} 
perturbative QCD corrections produce the following numerical  results 
\cite{Broadhurst:2004jx}
\begin{eqnarray}
I_G^{V}&=& \frac{1}{3}\bigg[
1+0.035521\bigg(\frac{\alpha_s}{\pi}\bigg)
  -0.58382\bigg(\frac{\alpha_s}{\pi}\bigg)^2\bigg]~~{ for }~~N_F=3\\
&=&\frac{1}{3}\bigg[ 1+0.038363\bigg(\frac{\alpha_s}{\pi}\bigg)
  -0.56479\bigg(\frac{\alpha_s}{\pi}\bigg)^2\bigg]~~{ for }~~N_F=4
\end{eqnarray}
which are higher  than the results of different analysis of NMC 
data \cite{Arneodo:1994sh}, though recently updated \cite{Abbate:2005ct}.
This discrepancy can be resolved by  
introducing to $I^{v}$ the  isospin-breaking 
effects in the Dirac sea $\overline{u}(x)<\overline{d}(x)$ as in Eq.(3.5).
It is interesting to note that within the large $N_c$-limit  
of the modified  soliton model of Ref. \cite{Witten:1983tx},
proposed in Ref. \cite{Diakonov:1996sr}, 
in the  {\bf planar approximation} one can get non-perturbative 
contribution of over $O(N_c^0)$  \cite{Pobylitsa:1998tk}
\begin{equation}
\frac{1}{2}(3I_G-1)=\int_0^1 dx \bigg(\overline{u}(x)-\overline{d}(x)\bigg)
=O(N_c^0)~~~~
\label{relation1}
\end{equation}
which is  absent in the perturbative series for $I_G^{v}$. 
Thus, the $O(N_c^0)$-effects, which do not manifest themselves in 
perturbative sector, may reflect the existence of 
flavour-asymmetry terms in the nucleon sea.
In the letter case the values of $I_G$ between 0.219 and 0.178 were obtained 
for a range of constituents quark mass between 350 and 420 ${\rm MeV}$,
in fair agreement with the announced NMC result $I_G^{exp}=0.235 \pm 
0.026$ at $Q^2=4$ GeV$^2$ \cite{Arneodo:1994sh}.

To summarize, we would like to emphasize that 
the bridge might exist between {\bf the $O(N_c^0)$ corrections}, 
which {\bf are absent} in the perturbative sector, but are manifesting 
themselves in the {\bf the non-perturbative regime}.

\section{Conclusions}
In this talk we would like to put extra attention on the importance 
of analytical multiloop analytical perturbative QCD calculations of 
characteristics of DIS. Indeed, the results of definite calculations 
reveal the existence of special topological structure in some 
physical quantities, related to phenomenology. Among them is the perturbative 
series for the Gottfried sum rule, which up to order $a_s^2$ corrections 
is defined by {\bf non-planar} $O(1/N_c^2)$ suppressed corrections only.
It is interesting to understand, whether this typical {\bf non-planarity} 
will continue in higher loops. The results of the talk of Ref. \cite{Rogal}
indicate that definite conjectures made in Ref. \cite{Broadhurst:2004jx} 
on continuation of {\bf non-planar}
suppression in the characteristics of $\nu N$ DIS  at the $a_s^3$ level 
remain true. The possible physical explanation  of this special 
structure of perturbative QCD series is the existence in the Dirac sea 
 of isospin symmetry 
breaking effect, which is generating light quark flavour asymmetry.
Another interesting physical feature of the similar origin is 
the perturbative generation 
of strange-antistrange asymmetry $s(x)-\overline{s}(x)$ in the nucleon sea 
\cite{Catani:2004nc}  
by  NNLO corrections $N_F({P}_{qq}^{s}- P_{q\overline{q}}^{s}) =
a_s^3 P_{NS}^{(2),s}$, where   $P_{NS}^{(2),s}$ term, calculated in 
Ref.\cite{Moch:2004pa} is proportional to $d^{abc}d_{abc}$ colour structure 
and result from {\bf ``light-by-light'' scattering} type diagrams.
Thus, one may expect, that the appearance of new colour structures 
in the perturbative series for DIS may be really connected to 
new physical conclusions, which should be studied more carefully.

\section{Acknowledgements}
I would like to acknowledge  constant interest to my works by my 
late colleague W. van Neerven, his 
constructive questions and comments, which started in 1990 
and unfortunately were interrupted after our e-mail discussion at the end of  
January,  2007. I am grateful to   
D.J.  Broadhurst and C.J. Maxwell for very productive collaboration
and to  S.I. Alekhin, S. Catani,  
J. Blumlein, A.V. Kotikov, 
S.A. Kulagin, G. Parente and C. Weiss for useful discussions, related   
to this report. It is the pleasure to thank the organizers
of XI International  Workshop in Advanced Computing and Analysis 
Techniques in Physics Research ACAT 2007, and in particular 
J.A.M. Vermaseren, for hospitality in Amsterdam. I  also 
acknowledge CERN Theory Unit for invitation and 
providing excellent working conditions.


\begin{thebibliography}{99}
\bibitem{Hatton} L.~ Hutton, {To what extent can we rely 
on the results of scientific calculations,} 
\emph{talk at 
ACAT07, Amsterdam, April 23, 2007}
\bibitem{Morhac} M.~ Morhac, {Error free algorithms to solve 
special and general discrete systems of linear 
equations,} \emph{talk at ACAT07, Amsterdam, 
 April 23, 2007}.
\bibitem{Beneke:1998ui}
  M.~Beneke,
\emph{Phys. Rept.} {\bf 317} (1999) 1.
\bibitem{Shirkov:2006gv}
  D.~V.~Shirkov and I.~L.~Solovtsov,
\emph{Theor. Math. Phys.} {\bf 150} (2007) 132.
\bibitem{Broadhurst:1998rz}
  D.~J.~Broadhurst,
 \emph{Eur.Phys. J.} {\bf C8} (1999) 311.
\bibitem{Broadhurst:2004jx}
  D.~J.~Broadhurst, A.~L.~Kataev and C.~J.~Maxwell,
 \emph{Phys. Lett.}  {\bf B590}  (2004) 76.
\bibitem{Catani:2004nc}
  S.~Catani, D.~de Florian, G.~Rodrigo and W.~Vogelsang,
 \emph{Phys.Rev.Lett.}  {\bf 93} (2004) 152003.
\bibitem{'tHooft:1973jz}
G.~'t Hooft,
\emph{Nucl. Phys} {\bf B72} (1974) 461.
\bibitem{Kataev:2001kk}
  A.~L.~Kataev, G.~Parente and A.~V.~Sidorov,
\emph{Phys.Part.Nucl.}  {\bf 34} (2003) 20
[\emph{Fiz. Elem. Chast. Atom. Yadra} {\bf 34} (2003) 43]
[{\tt hep-ph/0106221}].
\bibitem{Seligman:1997mc}
  W.~G.~Seligman {\it et al.},
\emph{ Phys. Rev. Lett.} {\bf 79} (1997) 1213.
\bibitem{Moch:2004pa}
S.~Moch, J.~A.~M.~Vermaseren and A.~Vogt,
\emph{Nucl.Phys}   {\bf B688} (2004) 101.
\bibitem{Parente:1994bf}
  G.~Parente, A.~V.~Kotikov and V.~G.~Krivokhizhin,
 \emph{Phys.Lett.}   {\bf B333} (1994) 190.
\bibitem{Kataev:1996vu}
  A.~L.~Kataev, A.~V.~Kotikov, G.~Parente and A.~V.~Sidorov,
\emph{Phys.Lett.} {\bf B388} (1996) 179.
\bibitem{Kataev:1997nc}
  A.~L.~Kataev, A.~V.~Kotikov, G.~Parente and A.~V.~Sidorov,
\emph{Phys.Lett.} {\bf B417} (1998) 374.
\bibitem{Santiago:1999pr}
  J.~Santiago and F.~J.~Yndurain,
\emph{Nucl.Phys.} {\bf B563} (1999) 45.
\bibitem{Kataev:1999bp}
  A.~L.~Kataev, G.~Parente and A.~V.~Sidorov,
\emph{Nucl.Phys.} {\bf B573} (2000) 405.
\bibitem{Kulagin:2000yw}
  S.~A.~Kulagin and A.~V.~Sidorov,
\emph{Eur.Phys. J.}  {\bf A9} (2000) 261
\bibitem{Larin:1993vu}
  S.~A.~Larin, T.~van Ritbergen and J.~A.~M.~Vermaseren,
\emph{Nucl.Phys.} {\bf B427} (1994) 41.
\bibitem{Larin:1996wd}
  S.~A.~Larin, P.~Nogueira, T.~van Ritbergen and J.~A.~M.~Vermaseren,
\emph{Nucl.Phys.} {\bf B492} (1997) 338.
\bibitem{Retey:2000nq}
  A.~Retey and J.~A.~M.~Vermaseren,
 \emph{Nucl.Phys.} {\bf B604} (2001) 281.
\bibitem{Blumlein:2004xt}
  J.~Blumlein and J.~A.~M.~Vermaseren,
\emph{Phys.Lett.} {\bf B606} (2005) 130.
\bibitem{Zijlstra:1992kj}
  E.~B.~Zijlstra and W.~L.~van Neerven,
\emph{Phys. Lett.} {\bf B297} (1992) 377.
\bibitem{Gorishnii:1985xm}
  S.~G.~Gorishny and S.~A.~Larin,
\emph{Phys.Lett.} {\bf B172} (1986) 109.
\bibitem{Moch:1999eb}
  S.~Moch and J.~A.~M.~Vermaseren,
\emph{Nucl.Phys.} {\bf B573} (2000) 853
\bibitem{Moch:2007gx}
  S.~Moch and M.~Rogal,
[{\tt arXiv:0704.1740 [hep-ph]}].
\bibitem{Larin:1991tj}
  S.~A.~Larin and J.~A.~M.~Vermaseren,
\emph{Phys.Lett.}  {\bf B259} (1991) 345.
\bibitem{van Neerven:1991nn}
  W.~L.~van Neerven and E.~B.~Zijlstra,
\emph{Phys.Lett}  {\bf B272} (1991) 127.
\bibitem{Zijlstra:1992qd}
  E.~B.~Zijlstra and W.~L.~van Neerven,
\emph{Nucl.Phys}   {\bf B383} (1992) 525.
\bibitem{Santiago:2001mh}
  J.~Santiago and F.~J.~Yndurain,
\emph{Nucl. Phys.}  {\bf B611} (2001) 447.
\bibitem{Maxwell:2002mt}
  C.~J.~Maxwell and A.~Mirjalili,
\emph{Nucl. Phys.} {\bf B645} (2002) 298.
\bibitem{Khorramian:2006wg}
  A.~N.~Khorramian and S.~Atashbar Tehrani,
\emph{JHEP} {\bf 0703} (2007) 051
\bibitem{Brooks:2006wh}
  P.~M.~Brooks and C.~J.~Maxwell,
 \emph{Nucl. Phys.} {\bf B} (2007) (in press);
[{\tt hep-ph/hep-ph/0610137}].
\bibitem{Alekhin:2005gq}
  S.~Alekhin,
\emph{JETP Lett.} {\bf 82} (2005) 628
[\emph{Pisma Zh. Eksp. Teor. Fiz.}  {\bf 82} (2005) 710].
\bibitem{Blumlein:2007dk}
  J.~Blumlein,
\emph{talk at the 15th Int. Workshop on DIS and Related Subjects (DIS2007),
Munich, Germany  April 17, 2007} 
[{\tt  arXiv:0706.2430 [hep-ph]}].
\bibitem{Bethke:2006ac}
  S.~Bethke,
\emph{Prog.Part.Nucl..Phys.}  {\bf 58} (2007) 351.
\bibitem{Martin:2007bv}
  A.~D.~Martin, W.~J.~Stirling, R.~S.~Thorne and G.~Watt,
[{\tt arXiv:0706.0459 [hep-ph]}].
 \bibitem{Kataev:2004wv}
   A.~L.~Kataev,
\emph{Phys.Part.Nucl.}  {\bf 36}  (2005) S168
[{\tt hep-ph/0412369}].
 \bibitem{Adler:1965ty}
 S.~L.~Adler,
\emph{Phys.Rev.}  {\bf 143}  (1966) 1114. 
 \bibitem{Dokshitzer:1995qm}
 Y.~L.~Dokshitzer, G.~Marchesini and B.~R.~Webber,
\emph{Nucl. Phys.}  {\bf B469}   (1996) 93. 
 \bibitem{Bjorken:1968dy}
 J.~D.~Bjorken,
\emph{ Phys.Rev.} {\bf 179} (1969) 1547.
 \bibitem{BVT}
 N.~N.~Bogolyubov, V.~S.~Vladimirov and A.~N.~Tavkhelidze, 
\emph{ Teor. Mat.  Fiz.} {\bf 12} (1972) 3;
 \emph{ibid}. {\bf 12}  (1972) 305.
 \bibitem{BSh}
 N. ~N. ~Bogolyubov and D.~V.~Shirkov, \emph{Introduction to the Theory 
 of Quantum Fields} Nauka, Moscow (1973, 1976, 1986).
\bibitem{Allasia:1985hw}
D.~Allasia {\it et al.},
\emph{ Z. Phys}  {\bf C28}   (1985) 312.
\bibitem{Kumano:1997cy}
 S.~Kumano,
\emph{ Phys.Rept.} {\bf 303} (1998) 183.
\bibitem{GP} 
G.~T.~Garvey and J.~C.~Peng,
\emph{ Prog. Part. Nucl. Phys.} {\bf 47}  (2001) 203.
\bibitem{ALK} 
A.~L.~Kataev, 
Proc. 11 Lomonosov Conference on Elementary Particle Physics,
Moscow, MSU, August 2003; ``Moscow 2003, Particle physics in laboratory,
space and universe'', World Scientific, p.194
[{\tt hep-ph/0311091}].
.
\bibitem{Arneodo:1994sh}
M.~Arneodo {\it et al.} [New Muon Collaboration],
\emph{ Phys. Rev.}  {\bf D 50}  (1994) 1.
\bibitem{Abbate:2005ct}
  R.~Abbate and S.~Forte,
\emph{Phys.Rev.} {\bf D72} (2005) 117503.
\bibitem{Gross:1973ju}
  D.~J.~Gross and F.~Wilczek,
\emph{Phys.Rev.} {\bf D8} (1973) 3633.
\bibitem{Ross:1978xk}
  D.~A.~Ross and C.~T.~Sachrajda,
\emph{Nucl.Phys.}{\bf B149} (1979) 497.
\bibitem{Curci:1980uw}
  G.~Curci, W.~Furmanski and R.~Petronzio,
\emph{Nucl.Phys.} {\bf B175} (1980) 27.
\bibitem{Borwein:1999js}
  J.~M.~Borwein, D.~M.~Bradley, D.~J.~Broadhurst and P.~Lisonek,
\emph{Trans. Am. Math. Soc}  {\bf 353} (2001) 907.
\bibitem{Kataev:2003en}
  A.~L.~Kataev and G.~Parente,
\emph{Phys.Lett.} {\bf B566} (2003) 120.
\bibitem{Kotikov:2005gr}
  A.~V.~Kotikov and V.~N.~Velizhanin,
 [{\tt hep-ph/0501274}].
\bibitem{Rogal}
M. Rogal, \emph{talk  at 
ACAT07, Amsterdam, April 26, 2007};\\ 
S. Moch, M. Rogal and  A. Vogt; DESY 07-048, 2007  (to appear).  
\bibitem{Witten:1983tx}
  E.~Witten,
\emph{Nucl.Phys.}{\bf B223} (1983) 433.
\bibitem{Diakonov:1996sr}
  D.~Diakonov, V.~Petrov, P.~Pobylitsa, M.~V.~Polyakov and C.~Weiss,
\emph{Nucl. Phys.} {\bf B480} (1996) 341.
\bibitem{Pobylitsa:1998tk}
  P.~V.~Pobylitsa, M.~V.~Polyakov, K.~Goeke, T.~Watabe and C.~Weiss,
 \emph{Phys.Rev.}  {\bf D59} (1999) 034024.








\end{thebibliography}
\end{document}